\documentclass[aps,prl,twocolumn,showpacs,groupedaddress]{revtex4}
\usepackage{graphicx}
\usepackage{dcolumn}
\usepackage{bm}
\usepackage{ae}
\usepackage{color}
\usepackage{aecompl}
\usepackage{amsmath}
\usepackage{amssymb}
\usepackage{epsfig}
\usepackage{multirow}
\begin{document}

\title{Colloidal gels assembled via a temporary interfacial scaffold}

\author{Eduardo Sanz, Kathryn A. White, Paul S. Clegg and  Michael E. Cates}
\affiliation
{SUPA, School of Physics and Astronomy, University of Edinburgh, Mayfield Road,
Edinburgh EH9 3JZ, UK}

\date{\today}

\begin{abstract}
The liquid-liquid phase separation of a binary solvent can be arrested by colloidal particles trapped at the interface
[K. Stratford {\it et al.}, Science \textbf{309}, 5744 (2005)]. We
show experimentally that the colloidal network so formed can remain stable after fully remixing the liquids, creating a new type of gel in which colloids in a single-phase solvent have locally planar coordination. We argue that  this structure is likely maintained by primary-minimum DLVO bonding of our charged colloids, created under strong compression by capillary forces. We present simulation evidence that the combination of a short-ranged attraction with a repulsive barrier 
can strongly stabilize such locally planar gels. 
\if{Our results exemplify how
liquid-liquid interfaces can be used as a temporary scaffolding during the
assembly of new colloidal materials.
}\fi

\end{abstract}

\maketitle

Understanding and controlling the assembly of colloidal 
particles is central to the creation of new soft materials. One way to induce a specific assembly route (leading to a target structure)
is to tailor the inter-particle interactions \cite{NM_2005_6_557}; this
strategy is assisted by continuing advances in colloidal synthesis \cite{NM_2005_6_557,othermoreexperimental}.
The structure of colloidal gels, for instance, depends crucially on the interaction
between the particles: spherically symmetric attractive \cite{N_2008_453_06931} or oppositely charged particles \cite{JPCB_2008_112_10861}
form low-density solids characterised by locally dense, amorphous, percolating branches. By contrast
particles interacting via short-ranged attraction plus long-ranged repulsion form gels whose branches have a 
Bernal-spiral structure \cite{PhysRevLett.94.208301,JPCB_2005_109_21942}. 
A second, distinct approach to get particles assembled is to guide them into a 
specific structure with the aid of a template, 
as is done using emulsion droplets as templates \cite{L_1996_12_2374,S_2002_298_1006} or by solvent demixing to form 
bijels \cite{S_2005_309_5744,NM_2007_6_966} , where the strong attachment energy of colloids to a
liquid-liquid interface is exploited. 
(Templating of molecules rather than colloids has also been successful, 
e.g., in the fabrication of porous silica monoliths via surfactant bilayer templates \cite{S_1997_25_552,JCP_1992_96_4174}.) 
A natural question is to what extent one can for colloids combine these two strategies 
-- interaction and template-based -- by tailoring interactions to maintain stability 
of a structure after deliberate removal of the template, which is then a temporary `scaffold'. 
This approach has been pursued to stabilise colloidal 
shells at the interface of emulsion droplets (`colloidosomes')\cite{S_2002_298_1006,L_2005_21_2693}. 
 
\textcolor{black}{In this Letter we follow this combined assembly strategy to create a low-density colloidal gel
 in a single-phase 
solvent that differs strongly from conventional, percolating branched gels \cite{N_2008_453_06931}}. 
The local structure of our gel comprises instead a web of locally 2D colloidal monolayers: 
for brevity, we call it a `monogel'. This has structural analogies with the fused silica monoliths 
of \cite{S_1997_25_552}. 
More pertinently, the structure is also that of the bijel
(bicontinuous interfacially jammed emulsion gel \cite{S_2005_309_5744}); 
the latter however contains a biphasic solvent.
The bijel is created by dispersing colloidal particles in a
binary-fluid mixture while it is in the single-fluid phase. When the
sample is quenched into the demixed region the liquids phase
separate. If the particles have neutral (or near-neutral)
wetting, they will sequester to the newly created
liquid-liquid interfaces; as the domains coarsen they will
jam, causing structural arrest. The ultimate shape of the domains reflects the demixing route:
for near-symmetric binary fluid pairs a bicontinuous domain arrangement is
pinned in place by the particle-laden interface \cite{NM_2007_6_966}. As so far conceived, the colloidal monolayer templated by this 
structure is held in place entirely by interfacial tension \cite{S_2005_309_5744}.
However, below we establish experimental conditions for which
this locally planar arrangement of the colloidal particles 
remains in place even after the liquids are remixed -- 
a situation stabilized, presumably, by attractive interactions. 
(At the low volume fractions used, repulsive colloidal particles 
would be uniformly dispersed in equilibrium.) Thus we obtain a  
bonded gel in a single solvent, with locally monolayer organization. 
We then show, using computer simulations, that quite simple combinations 
of attractive and repulsive interactions are indeed capable of stabilising 
this `monogel' state. Whether the experimental interactions are this 
simple is discussed in the final part of the Letter.

\begin{figure}
\includegraphics[clip,width=0.5\textwidth,angle=-0]{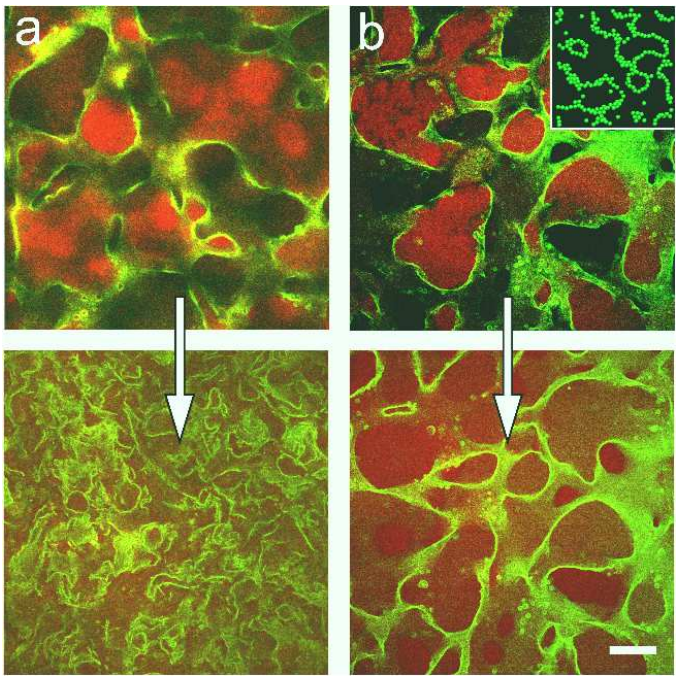}
\caption{\small (Color online) Confocal microscopy images of critical-composition water(dark grey (black))-lutidine 
(grey (red)) mixtures containing silica particles (bright grey (green)).  Scale bar is 100 $\mu m$.
The upper images were taken with the samples at 40$^{\circ}$C and the lower images were taken after cooling 
into the single-fluid phase. Rhodamine B dye (red) shows the presence of a lutidine-rich phase. 
In (a) the bijel was cooled sooner than in (b), 
where the colloidal gel remained stable after the liquid phases had remixed. Further particle details are given in
\cite{JPCM_2008_20_494223}. See EPAPS Document No. [] (www.aip.org/pubservs/epaps.html) 
for movies of both experiments. 
Inset to (b): a section from the starting bijel colloidal network used in our computer simulations.}
\label{experimental}
\end{figure}

We prepared  water--2,6-lutidine bijels stabilized 
by colloidal particles ($\sim$0.4\,$\mu$m diameter St\"{o}ber silica)
using the formulation route described in Ref.\cite{JPCM_2008_20_494223}. The system is visualized using 
a confocal microscope; particles are dyed with fluorescein isothiocyanate (green) and the lutidine-rich domain with rhodamine B (red) \cite{JPCM_2008_20_494223}. 
\textcolor{black}{
In the experiments we report here (Fig.~\ref{experimental})  the newly formed bijel is aged for some time and subsequently
cooled to a temperature where the liquids return to the single-fluid (mixed) phase. During remixing the rhodamine B dye 
diffuses throughout the sample. Our results differ depending on the aging time (Fig.~\ref{experimental}). 
For aging times less than $\sim$ 30 minutes (Fig.~\ref{experimental}(a)) the structure disintegrates as the fluids remix,  as expected.
By contrast, if the bijel is aged for {\em ca.} a day  (Fig.~\ref{experimental}(b)) the colloidal ``skeleton" 
remains surprisingly stable even after the liquids have fully remixed and all fluid
interfaces have disappeared: Fig.~\ref{experimental}(b)-bottom is thus a monogel.} 

To understand our experiments we need to know what sort of inter-particle 
interactions are capable of 
stabilizing a monogel network after the interfacial `scaffolding' is removed. To address this question
we carry out computer simulations monitoring the evolution of an initial monogel arrangement
for different interactions. 
We take the initial particle configuration from a 0.5M timestep Lattice-Boltzmann bijel study
in which both fluids and the particles are simulated \cite{L_2008_24_6549}. This algorithm is prohibitively expensive for the long-time regime addressed here; therefore, from that starting point, we run $NVT$ Monte Carlo (MC) simulations. In the limit of small particle displacements, $\Delta x_{max}$, these
mimic the Brownian motion
of particles suspended in a homogeneous fluid \cite{CPL_1991_185_335}; 
here we set $\Delta x_{max} \equiv$ 2\% of the particle diameter $\sigma$.
An MC cycle (a trial move per particle) is roughly equivalent to
$(\Delta x_{max})^2\cdot a/6$ Brownian times, $t$, where $a$ is the acceptance 
rate of the displacement moves \cite{edudavide}.
The starting
bijel network contains 8239 particles occupying 20\% of the volume.  
Figure \ref{experimental}(b) (inset)
shows a snapshot of a 2-D cut of the initial configuration.
We have used two simple interaction potentials to try to stabilize the 
monogel: a short-ranged attraction (SRA), meant to  act as a `glue' between the particles, 
and a short-ranged attraction plus a long-ranged repulsion (SRA-LRR), 
tailored to kinetically stabilize colloidal monolayers (such as the monogel walls) by
opposing the approach of second neighbors (and beyond) to become nearest neighbors:
\begin{equation}
u_{SRA}(r) =
\begin{cases}
\infty & r \le \sigma\\
-\epsilon & \sigma < r \le \lambda \sigma\\
0  & r > \lambda \sigma\\
\end{cases} 
\end{equation}
\begin{equation}
u_{SRA-LRR}(r) =
\begin{cases}
\infty & r \le \sigma\\
-\epsilon & \sigma < r \le \lambda \sigma\\
\frac{2\epsilon}{3(2-\lambda)}\left(1-\frac{r}{2\sigma}\right) &  \lambda \sigma < r \le 2\sigma\\
0 & r > 2\sigma 
\end{cases}, 
\label{ramp}
\end{equation}
Here $\epsilon$ is the attractive well depth, and $\lambda$ the
range of the attraction.  We have used $\epsilon = 10 k_BT$ and $\lambda$ ranging from 1.025 to 1.5. 
The selected SRA potential is the simplest possible, a square well;
the SRA-LRR combines this at short distances, $r$, with a repulsive barrier 
of height $\epsilon/3$ at $r=\lambda\sigma$ ramping down to zero at $r=2\sigma$ (see Fig.~\ref{prim_sec}(a) and (b)). 
Compared to a DLVO potential (see below) the choice of potential (2) allows one to study unambiguously the 
competition between SRA and LRR.

\begin{figure} 
\includegraphics[clip,width=0.3\textwidth,angle=-0]{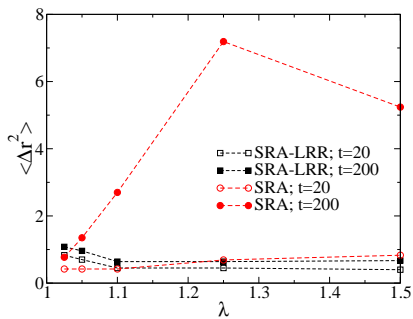}
\caption{\small 
(Color online) Mean square displacement at two different times for SRA and SRA-LRR interactions. 
Particles interacting via SRA-LRR barely move, their displacement being clearly subdiffusive. 
}
\label{msd}
\end{figure}

\begin{figure} 
\includegraphics[clip,width=0.4\textwidth,angle=-0]{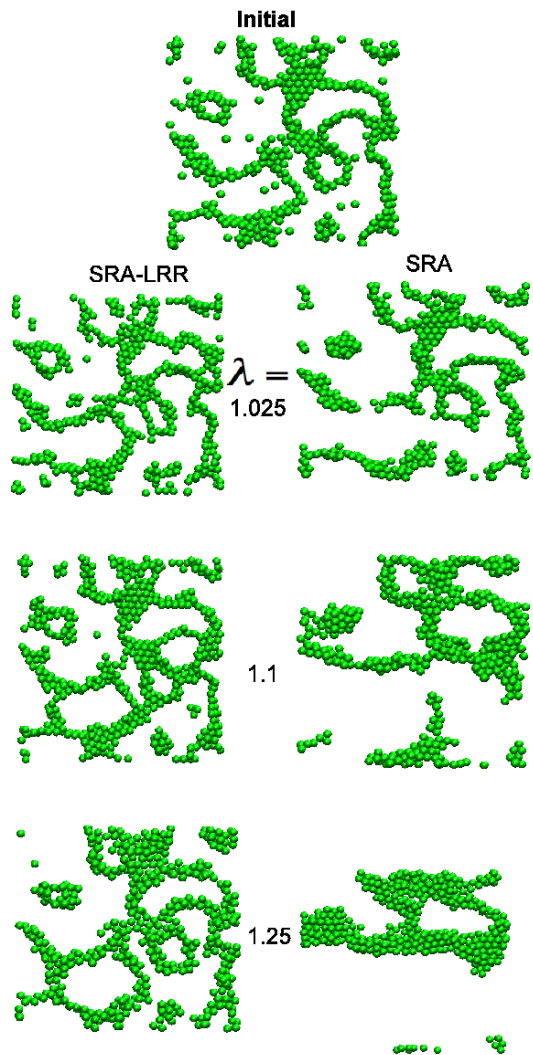}
\caption{\small (Color online) A slab of the initial configuration is compared to those obtained after simulating 200 t
for both SRA-LRR and SRA interactions and different attractive ranges $\lambda$.} 
\label{structure}
\end{figure}

Fig.~\ref{msd} shows how effective these potentials are in preventing `melting' of the monogel as monitored by particle displacement. 
In addition, a visual comparison of the initial and the final structures is provided in Fig.\ref{structure}. 
Clearly, SRA-LRR is very effective at trapping the target structure. 
Only for extremely short ranges can SRA alone 
sufficiently reduce Brownian motion within the monogel to 
effectively quench its structure (likely also requiring very deep 
wells to achieve this). Moreover, without either directional forces 
or a local barrier to rearrangement, any finite piece of 
monolayer can roll up from the edges to gain the increased 
bonding energy of a locally 3D structure, so that stabilization 
of the monogel solely by SRA would be critically dependent 
on monolayer continuity. 
By contrast, when a long-ranged repulsion is added, particles have to 
surmount an additional potential barrier to locally rearrange from 2D to 3D coordination, and planar structures
remain stable to much larger values of $\lambda$.
For finite barrier heights some slow evolution is of course inevitable and in simulations we choose a modest one ($10k_BT/3$) to allow reasonable run times. Crucially however, in the laboratory much larger barriers can be introduced {\em without} impeding the initial templating step
via the bijel route. This is because the interfacial energies that jams particles into contact lie much further still above
the thermal scale \cite{S_2005_309_5744}.

\begin{figure}
\includegraphics[clip,width=0.5\textwidth,angle=-0]{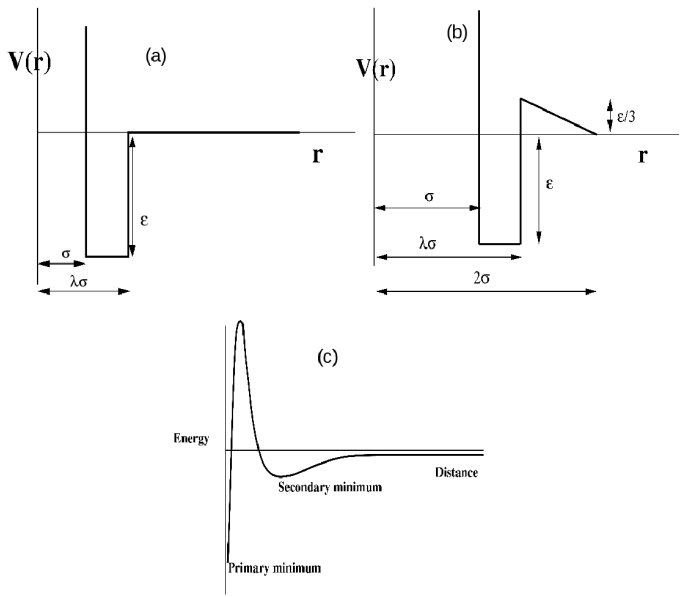}
\caption{\small (a) A square-well attractive interaction (SRA). 
(b) A square-well attraction combined with a ramp repulsion for longer distances (SRA-LRR). 
(c) Sketch of DLVO potential \cite{Israelachvili}. 
}
\label{prim_sec}
\end{figure}

Let us discuss now the nature of the colloidal interactions in the experiment of Fig.~\ref{experimental}. 
Bond formation occurs in the bijel template with particles trapped at liquid-liquid interfaces, 
a geometry in which the colloidal interactions compete with interfacial tension and, due to the 
heterogeneous solvent, cannot be described precisely. However they should not be very different 
from those acting in the 
final monogel which contains a single `averaged' solvent. These interactions 
can be modelled by a standard DLVO form with a van der Waals attraction at short 
distances and a screened electrostatic repulsion (caused by deprotonation 
of surface silanol groups) at larger ones.  There are then two possible 
sources of attraction (see Fig.\ref{prim_sec}(c) for a sketch): 
a deep primary minimum at short distances, and a weaker secondary 
minimum at larger ones \cite{Israelachvili}. 
It is highly implausible that the (long-ranged) secondary minimum is 
the origin of monogel stability; our simulation results show that to achieve stability we need much closer contacts, 
or a barrier, or preferably both. To access the primary minimum, however, the 
DLVO (Coulomb) barrier must be crossed. Our particles are stable against 
aggregation when suspended in the single phase solvent prior to bijel formation, 
so this barrier is clearly large compared to thermal energies. 
Nonetheless, capillary forces present during the bijel templating step could overcome this barrier. 
A tentative estimate of the maximum repulsive force between our particles using a DLVO potential
yields $(1.5 \pm 0.5) \times 10^4 k_BT/\sigma$ (the estimated interaction parameters are \cite{JPCM_2008_20_494223}: 
surface charge density $0.01 e/$nm$^{-2}$, Debye length $10$ nm, 
Bjerrum length $1.02$ nm, and Hamaker constant $3.6\times 10^{-21}$ J). On the other hand, 
we can estimate the attractive capillary force as $\simeq\gamma \sigma\tan(\pi/6)$ (for particles in close contact 
in a hexagonal raft geometry). This gives ~ $2\cdot10^4 k_BT/\sigma$ 
(with an interfacial tension in the demixed fluids at 40$^o$C of $\gamma \simeq 0.22$ mN m$^{-1}$ \cite{JCED_1993_38_516}). 
Although our estimates are quite rough, both forces have the same order of magnitude, making plausible the scenario in which
the monogel forms as a consequence of capillary attractions overcoming electrostatic repulsion. 

A remaining mystery is why the bijel has to be aged for the monogel network 
to remain stable after remixing the fluids. One possibility is that the capillary force 
does not quite remove the barrier but leaves it low enough to be crossed thermally 
on this timescale. Since both the capillary and the DLVO energy 
scales are large compared to $k_BT$, such close cancellation is {\it a priori} improbable. 
\textcolor{black}{ However,
it seems quite likely that the capillary forces 
increase significantly during aging, as particles rearrange and the number of close contacts increases
\cite{L_2008_24_6549}. 
In this case the incubation time would correspond to the time it takes an evolving
barrier to be  reduced to a level where thermal crossing is possible.
Once in the primary minimum, covalent bonds 
may develop between surface silanol groups \cite{iler}, helping to lock in the monogel 
structure.  But, given the depth of the DLVO primary 
minimum, it would be surprising if covalent bonds were necessary.}

According to our DLVO estimates, the repulsive barrier in the experiments is not 
only far higher but also much shorter ranged than in the simulations. 
However, the underlying cause for monogel stability is unchanged: 
short-range bonds complemented by a repulsion for longer distances that hinders rearrangements.  
Moreover, the simulations predict that a monogel structure can be stabilized
for a wide range of parameters within a SRA-LRR interaction. 
Accordingly, other than by the DLVO potential a monogel structure could perhaps also be stabilized by combining 
a (weakly screened) Coulomb repulsion with attractions from
non-adsorbing polymers \cite{PhysRevLett.94.208301} or
DNA linkages \cite{N_1996_382_607}, or by rotating magnetic fields \cite{juretobe}. 
This does not rule out the possibility that other types of interaction ({\it e.g.}, orientational bonding rather than a SRA-LRR) 
could also stabilise the monogel. 

In conclusion, we have shown by experiment that a colloidal monolayer network 
templated by liquid-liquid interfaces
can remain stable after the fluids remix and the interfaces disappear, giving rise to a new sort of gel we call a `monogel'.  
Thus the interface acts as a scaffold which can later be removed, if the colloidal interactions are well chosen.
Our computer simulations establish that a combined short-ranged attraction and long-ranged repulsion 
can efficiently maintain such locally planar geometry. 
The monogel structure and formation route are distinct from those seen in conventional 
colloidal gels \cite{JPCM_2007_19_323101}. We have proposed, as a 
likely explanation for stability in our experiments, that during templating the colloids are 
forced into the primary attractive well of a DLVO potential, by the action of capillary forces. 
We plan to test this hypothesis by studying the behavior of the particles at flat liquid-liquid 
interfaces with the aid of a Langmuir trough. 
By contrast with conventional gels, if a monogel is fluidized by strong shearing the colloids should not re-aggregate. This fundamental
difference might be exploited for practical purposes.
Moreover, thermal cycling of such a system into the 
demixing region could  rejuvenate the structure, and perhaps provide new avenues 
to the control of gel rheology.

{\it Acknowledgements:} We thank E Kim for the initial 
bijel coordinates, J Tavacoli and J H J Thijssen for discussions and EPSRC GR/E030173 
and SE/POC/8-CHM-002 for financial support. MEC is funded by the Royal Society and ES by a Marie Curie EIF. 
The simulations were run using Edinburgh Compute and Data Facility. 
\bibliographystyle{apsrev}


\end{document}